\newcommand{\ket}[1]{|\,#1\,\rangle}
\newcommand{\bra}[1]{\langle\,#1\,|}

\documentclass[12pt]{article}  
\usepackage{fullpage} 
\title{Drift-Diffusion in Mangled Worlds Quantum Mechanics}
\author{Robin Hanson 
\\  
George Mason University\thanks{\emph{rhanson@gmu.edu} 
http://hanson.gmu.edu 703-993-2326 FAX: 703-993-2323
MSN 1D3, Carow Hall, Fairfax VA 22030}
}

\date{March 17, 2003.}

\begin{document}

\maketitle


\begin{abstract}
In Everett's many worlds interpretation, where quantum measurements are seen as decoherence events, inexact decoherence may let large worlds mangle the memories of observers in small worlds, creating a cutoff in observable world size.  I solve a growth-drift-diffusion-absorption model of such a mangled worlds scenario, and show that it reproduces the Born probability rule closely, though not exactly.  Thus deviations from exact decoherence can allow the Born rule to be derived in a many worlds approach via world counting, using a finite number of worlds and no new fundamental physics.
\end{abstract}



\subsection{Introduction}

Traditional quantum mechanics suffers from many ambiguities regarding quantum measurements.  Many worlds approaches try to resolve these ambiguities by identifying measurements with decoherence events produced by standard linear quantum evolution \cite{everett:manyworlds,dewittgraham:manyworlds}.  In such events, off-diagonal density matrix elements are often naturally and dramatically suppressed due to coupling with a large environment \cite{dowkerhalliwell:decoherence}.  

Unfortunately, the many worlds approach still suffers from the problem that, when there are a finite number of worlds, the straightforward way to calculate probabilities, i.e., counting the fraction of worlds with a given outcome, does not produce the standard Born probability rule \cite{kent:againstmanyworlds,auletta:interpqm}.  Some have tried to resolve this by adding new fundamental physics, such as non-linear dynamics \cite{weissman:measureprob}, or an infinity of worlds per quantum state which diverge via an unknown process \cite{albertloewer:manymindsinterpretation}.   Others propose new decision theory axioms saying we do not care about the number of worlds that see an outcome \cite{deutsch:quantumprobfromdecisions,wallace:quantumdecisiontheory}.  

A mangled worlds variation proposes to resolve the Born rule problem using only assumptions that are about the behavior of standard linear quantum evolution, assumptions that can in principle be checked by theoretical analysis of common quantum systems.  The idea is that decoherence is never exact, and so while decoherence makes off-diagonal terms small relative to a large world, they can be large relative to a small enough world.  This may plausibly mangle very small worlds, either destroying the observers in such worlds, or changing them into observers who remember events from large worlds \cite{hanson:worldhit}.  

This paper presents an explicit growth-drift-diffusion-absorption model of such a mangled worlds scenario.  It assumes that mangling is a sudden irreversible process, that typical experimental tests of the Born rule include many uncounted background decoherence events, and that the region in world size where worlds become mangled is relatively narrow and located near the median of the distribution of measure.  Closed form expressions are given showing that the Born rule is followed closely, though not exactly, in this model.  Thus most unmangled worlds would remember having observed frequencies very near that predicted by the Born rule, even though in fact Born frequencies do not apply to the vast majority of worlds, and even though such frequencies are not observed in the very largest worlds.  

\subsection{Drift-Diffusion Of All Worlds}

Consider a single initial unit-magnitude world that repeatedly undergoes decoherence events $e$.  (And for now ignore world mangling.)  During each event $e$, each pre-existing world $i$ splits into a set $J(i,e)$ of resulting worlds $j$, each of which gets some fraction $F_{j i e}$ of the original world's measure.   That is, if $m_i$ is the measure (or size) of world $i$, then $m_j = F_{j i e} m_i$, where$\sum_{j \in J(i,e)} F_{j i e} = 1$.  Assume that these fractions do not depend on the particular world being split, so 
that $F_{j i e} = F_{j e}$ and $J(i,e) = J(e)$.  If so, then after there have been enough events so that the variance of each set $\{\ln(F_{je})\}_{j \in J(e)}$ is small compared to this summed over $e$, the central limit theorem of statistics ensures that the resulting set of worlds will approach a log-normal distribution over measure $m$.  

In terms of log size $x = \ln(m)$, the distribution of worlds would be normal with some mean $\tilde{x}<0$ and standard deviation $\sigma > 0$.  Since the total measure of all worlds is conserved, the total number of worlds must be $e^{-\tilde{x}-\sigma^2/2}$.  If there were a constant rate of decoherence events, so that $\tilde{x} = -vt$ and $\sigma^2 = w t$, then the distribution of worlds would be 

\begin{equation}
\mu_0(x,t) = \frac{1}{\sqrt{2\pi} \sigma} \exp((v-\frac{w}{2})t - \frac{(x-vt)^2}{2 \sigma^2}) .
\label{normal} \end{equation}
The measure held by these worlds would also be normally distributed over $x$, with the same standard deviation $\sigma$, but a higher mean of $\hat{x} = \tilde{x} + \sigma^2$.  These values $\tilde{x}$ and $\hat{x}$ are also the positions of the median-sized world and the median measure, respectively.

For example, after $N$ binary decoherence events where the two possible outcomes have relative measure $p > 1/2$ and $1-p$, there would be $2^N$ worlds, with median measure and standard deviation given by
\[ \hat{x} = N \hat{x}_1 = N(p\ln(p)+(1-p)\ln(1-p)) , \]
\[ \sigma = \sqrt{N} \sigma_1 = \sqrt{N}  \sqrt{p(1-p)} \ln(\frac{p}{1-p}) , \]
with $\tilde{x}_1 = \hat{x}_1 - \sigma_1^2$.  A constant rate $r$ of such events would give $v = - r \tilde{x}_1$ and $w = r 
\sigma_1^2$.

The distribution $\mu_0(x,t)$ of equation~\ref{normal} solves the linear growth-drift-diffusion equation

\begin{equation}   
\dot{\mu} = v(\nabla \mu + \mu) + (w/2)( \nabla^2 \mu - \mu) .  
\label{diffuse} \end{equation}
for $t>0$ given the delta function initial condition 

\begin{equation}   
 \mu(x,0) = \delta(x) . \
\label{initial} \end{equation}

\subsection{The Mangling of Worlds}

Decoherence events quickly suppress the magnitude of off diagonal density matrix terms relative to diagonal terms. Specifically, given two worlds, large $L$ and small $s$, we have 
\[ | \rho_{Ls} |^2 \leq  \epsilon^2(t)  | \rho_{LL} | | \rho_{ss} |   , \]
where $|\rho_{ab}| \equiv \bra{a}\rho_{ab}\ket{b}$, and coherence $\epsilon(t)$ typically falls at a rapid exponential rate for a great many doubling times \cite{dowkerhalliwell:decoherence}.  However, in the models that have been solved so far,  coherence $\epsilon(t)$ typically eventually asymptotes to a small but non-zero level \cite{dowkerhalliwell:decoherence,dowkerhalliwell:decoherencesummary,namikietal:decoherence,unruhzurek:quantumbrownianmotion}.

As we have seen in the previous section, the relative magnitude between any two random worlds increases as $e^{\sigma} = e^{\sqrt{wt}}$.  As discussed elsewhere \cite{hanson:worldhit}, if $\epsilon(t)$ eventually falls slower than this rate, then eventually the off diagonal terms $\rho_{sL}$ and $\rho_{Ls}$ must greatly influence the evolution of the smaller diagonal term $\rho_{ss}$, even though these off diagonal terms have little influence on and are mainly driven by the larger diagonal term $\rho_{LL}$.  Thus the evolution of smaller worlds is eventually driven by larger worlds, plausibly ``mangling" those smaller worlds, i.e., either destroying their observers or turning them into observers who remember outcomes from a large world.  Remaining observers would thus only remember the histories of unmangled worlds.  

Let us conjecture that the mangling of worlds is a sudden, irreversible, and global process.  Specifically, let us assume that there is a mangling region in world size, so that a world that has always remained larger than this region remains unmangled, and that a world that was once smaller than this region became suddenly and forevermore mangled.  Let us also conjecture that the relevant standard deviation $\sigma$ is the result of many decoherence events, most of which are not counted in experiment tests.  This suggests that the mangling region is relatively narrow compared to $\sigma$.  Finally, let us assume that since it is the measure of some worlds that mangles other worlds, this mangling region remains close to the median measure $\hat{x}$ that would describe the distribution of all worlds were there no mangling.

If unmangled worlds evolve locally just as all worlds would were there no mangling, and if initially all worlds are unmangled, then the distribution $\mu_1(x,t)$ of unmangled worlds should satisfy equations~\ref{diffuse} and~\ref{initial}, just as the distribution $\mu_0(x,t)$ of all worlds under the no mangling assumption.   To model our assumption of a mangling region narrow compared to $\sigma$ and remaining close to the median measure $\hat{x} = -(v-w)t$, let us impose on the unmangled distribution $\mu_1(x,t)$ the additional boundary condition 

\begin{equation}   
\mu(x_b(t),t) = 0 ,
\label{absorb} \end{equation}
for all $t \geq 0$, where $x_b(t) = \hat{x}-\epsilon$, $\epsilon > 0$, and we limit our attention to $x \geq x_b(t)$.  This is an absorbing boundary condition, which says that every world which reaches the point $x_b$ from above immediately falls out of the distribution of unmangled worlds.   

\subsection{Solving the Drift-Diffusion Model}

To solve the set of equations~\ref{diffuse},~\ref{initial}, and~\ref{absorb}, let us first factor out the common exponential growth via $\mu(x,t) = \nu(x,t) e^{(v-(w/2))t}$, so that the diffusion equation~(\ref{diffuse}) becomes

\[  \dot{\nu} = v \nabla \nu + (w/2) \nabla^2 \nu , \]
for $\nu(x,t)$.  Next, let us transform from $x$ to a coordinate $y=x-x_b(t)$ that moves along with the absorbing boundary, so that equations~\ref{diffuse},~\ref{initial}, and~\ref{absorb} become
\begin{eqnarray*} 
 \dot{\nu}(y,t) &=& w \nabla \nu(y,t) + (w/2) \nabla^2 \nu(y,t) ,
\label{diffuse1}  \\
 \nu(y,0) &=& \delta(y-\epsilon) ,
\label{initial1}  \\
\nu(0,t) &=& 0 .
\label{absorb1} 
\end{eqnarray*}
Finally, use dimensionless units $z = 2y$ and $s = wt$, so that for $z,s \geq 0$ the equations are 
\begin{eqnarray} 
 \dot{\nu}(z,s) &=&  \nabla \nu(z,s) + \nabla^2 \nu(z,s) ,
\label{diffuse2}  \\
 \nu(z,0) &=& \delta(z-2\epsilon) ,
\label{initial2}  \\
\nu(0,s) &=& 0 .
\label{absorb2} 
\end{eqnarray}
Fortunately, others have already solved a closely related set of equations for $s \geq 0, z \in [0,L],$
\begin{eqnarray} 
 \dot{\nu}(z,s) &=&  \nabla \nu(z,s) - \nabla^2 \nu(z,s) ,
\label{diffuse3} \nonumber \\
 \nu(z,0) &=& f_0(z) ,
\label{initial3}  \\
\nu(0,s) &=& \nu(L,s) = 0 ,
\label{absorb3}  \nonumber
\end{eqnarray}
for drift diffusion between two absorbing barriers \cite{farkasfulop:1Ddriftdiffuseabsorb}.  
If we take $z \rightarrow L-z$ and $L\rightarrow \infty$, the solutions of these equations can be transformed into solutions of equations~\ref{diffuse2} and ~\ref{absorb2}, and of equation \ref{initial3}, which generalizes equation~\ref{initial2}. 
These solutions are
\[ \nu(z,s) = \frac{2}{\pi} e^{-s/4-z/2} \int_0^\infty g_0(k) e^{-k^2 s} \sin(kz)  dk  \]
\[ g_0(k) = \int_0^\infty f_0(z) e^{z/2} \sin(kz) dz .\]
For the specific initial condition of equation~\ref{initial2}, i.e., $f_0(z) = \delta(z-2\epsilon)$, this becomes
\[ \nu(z,s) = \frac{e^{\epsilon - s/4 - z/2}}{\sqrt{4 \pi s}}  \left( \exp(-\frac{(z-2\epsilon)^2}{4s}) - \exp(-\frac{(z+2\epsilon)^2}{4s}) \right) . \] 
If we return this to dimension-full units, put back in the exponential growth, but retain the boundary relative coordinate $y= x-x_b(t)$, we get solutions to our original equations of interest, i.e., equations~\ref{diffuse},~\ref{initial}, and~\ref{absorb}, except in terms of $y$ instead of $x$.  We find that an initial distribution $\mu_1(y,0) = \delta(y-\epsilon)$ of unmangled worlds evolves into

\[ \mu_1(y,t;\epsilon) = \sqrt{\frac{\pi}{8 w t}} e^{\epsilon - y + (v-w) t} \left(\exp(-\frac{(y-\epsilon)^2}{2wt}) - \exp(-\frac{(y+\epsilon)^2}{2wt}) \right) . \] 
For $\epsilon \ll \sqrt{wt}$, a good approximation to this is
\[ \mu_1(y,t;\epsilon) = \frac{\epsilon e^{\epsilon}}{\sqrt{2\pi}} \frac{e^{(v-w) t}}{(wt)^{3/2}} y \exp(-y - \frac{y^2}{2wt}) . \]
We use this approximation from here on. 
It integrates to give a total unmangled world count 

\[  W(t;\epsilon) \equiv \int_0^\infty \mu(y,t;\epsilon) dy 
  = \frac{\epsilon e^{\epsilon}}{2}  e^{(v-w) t}  \left[ \sqrt{\frac{2}{\pi w t}}  -  e^{wt/2} \mbox{erfc}\left(\sqrt{\frac{wt}{2}}\right) \right]  . \]
Note that if $v>w$ then the total number of unmangled worlds, which grows as $e^{(v-w) t}$, will increase with time,  
even though it becomes an exponentially decreasing fraction of the number of all worlds, which grows as $e^{(v-(w/2))t}$.  To predict our existence in an ummangled world, the mangled worlds approach must predict that $v > w$.  

\subsection{Born Rule Accuracy}

How well does this diffusion process reproduce the Born rule?  Let each distinct measurement outcome $k$ be associated with a set of $G_k > 1$ child worlds per parent world, each of which is a factor $F_k < 1$ smaller than its parent.  The Born rule says that the probability of an outcome is given by $F_k G_k$.  To see how closely does this mangled worlds approach  conforms with the Born rule, let us calculate the number of unmangled worlds as associated with such an experimental outcome.  

Specifically, starting at time $t=0$ let a single unmangled world of size $y=\epsilon$ (relative to the mangling region at $y=0$) evolve into a distribution $\mu(y,t_1;\epsilon)$ at time $t_1$, where $1 \ll w t_1 \gg \epsilon$.  This is not intended to be the distribution of all worlds, but rather the distribution of all worlds consistent with the initial conditions of a given experiment to test the Born rule.  It is the result of decoherence events both during and before the experiment, events that are not counted in the statistics of the experiment. Given what we know about the experiment, we do not know which of these worlds we are in, and so we must average over these worlds when making experimental predictions.  

At time $t_1$, let each world with value $y_1$ in $\mu(y,t_1;\epsilon)$ be split into $G \geq 1$ worlds, each of which is a factor $F \leq 1$ smaller, so that it has the value $y = y_1 + \ln(F)$.  Let each of these worlds then evolve to produce more worlds over a longer time period $t_2 \gg t_1$.  For the Born rule to apply exactly, the number of unmangled worlds in the final distribution at time $t_2$ should go as $FG$.  (It would not contradict observations if the Born rule were violated soon after $t_1$, in worlds that were not mangled but were about to become mangled.  What we actually observe are long-existing historical records of experiments testing the Born rule.) 

With a little help from Mathematica, the final unmangled world count is found to be 
\begin{eqnarray*}
\lambda(F,G;t_1,t_2,\epsilon) &\equiv& G \int_0^\infty W(t_2;y) \mu(y-\ln(F),t_1;\epsilon) dy  \\
   &=& FG \, \mbox{erfc}(\frac{-\ln(F)}{\sqrt{2w t_1}}) \frac{\epsilon e^{\epsilon}}{4} e^{(v-w)(t_1+t_2)}  \left[ \sqrt{\frac{2}{\pi w t_2}} - e^{w t_2/2}  \mbox{erfc}\left( \sqrt{\frac{w t_2}{2}} \right) \right] . 
\end{eqnarray*}
The key thing to notice here is that when $w t_1$ is large, the Born rule correction

\[ \gamma(F,G) \equiv \frac{1}{FG} \frac{\lambda(F,G;t_1,t_2,\epsilon)}{\lambda(1,1;t_1,t_2,\epsilon)} = \mbox{erfc}(\frac{-\ln(F)}{\sqrt{2w t_1}}) .\]
changes \emph{very} slowly in the factor $F$.  For example, when $w t_1 = 10^{10}$, it requires a factor of $F=e^{-10^5} < 10^{-43000}$ to get the relative number of worlds to be $\gamma(F,G) \approx 1/3$.  
Thus this approach is very nearly consistent with the Born rule, while leaving open the possibility of small experimentally detectable deviations from the Born rule.  

\subsection{Conclusion}

This paper has presented an explicit growth-drift-diffusion-absorption model of a mangled worlds scenario.  Under this scenario, inexact decoherence results in larger worlds suddenly and irreversibly mangling any worlds that reach a narrow region in world size.  Closed form expressions are given showing that this model reproduces the Born rule closely, but not exactly.  Thus most unmangled observers would remember having observed nearly Born frequencies, allowing us to reconcile the many worlds approach with the Born rule without invoking new fundamental physics or decision theory axioms.  

\bibliographystyle{aps} 
\bibliography{bib}  

\end{document}